\documentclass{jetpl}

\twocolumn
\lat

\title{Two-level systems and mass deficit in quantum solids}

\rtitle{Two-level systems and mass deficit in quantum solids}

\sodtitle{Two-level systems and mass deficit in quantum solids}

\author{S.\,E.\,\,Korshunov\/\thanks{e-mail: serkor@itp.ac.ru}}

\rauthor{S.\,E.\,\,Korshunov}

\sodauthor{Korshunov}

\address{L.\,D.\,\,Landau Institute for Theoretical Physics,
Kosygina 2, Moscow 119334, Russia}

\dates{25 June 2009}{*}

\abstract{We study a disordered quantum solid incorporating two-level
systems in which a group of atoms (or a single atom) can experience
coherent tunnelling between two different positions and demonstrate
that an effective mass deficit induced by the presence of
such objects can manifest itself only at relatively high frequencies
and should vanish in the low-frequency limit. The crossover to the
regime which can be associated with the appearance of an effective mass
deficit has been observed in recent torsional oscillator experiments.}


\begin{document}
\maketitle
\vspace*{-79.2mm}
\hspace*{19mm}\makebox{ {\bf 90},  167-170 (25 July 2009)}
\vspace*{70.6mm}

{\bf 1. Introduction.} After a number of torsional oscillator experiments
[1\ch 3] demonstrated that at low temperatures solid $^4$He
behaves itself as if it contained a superfluid component which remains at
rest when an external force is applied to the sample,
{an idea was put forward}
that these properties can be explained by the presence of two-level
systems (TLSs) in which a group of atoms (or a single atom) experiences
{classical \cite{BGNT} or quantum \cite{A07,A08}} tunnelling
between two localized positions shifted with respect to each other. To
some extent this conjecture is confirmed by the dependence of the
experimentally observed mass deficit on the degree of disorder in the
\makebox{solid \cite{RR}.}

However, the analysis of {Refs. \cite{A07,A08}} has led to a rather
paradoxical conclusion that the effective mass deficit induced by TLSs
should remain non-vanishing
even in the adiabatic limit including the really stationary situation
of the thermodynamic equilibrium, when experimentally
the system behaves as if $\rho_{s}=0$ \cite{DHB}.
The origin of this conclusion can be traced to the assumption that the
bulk velocity of a solid is a classical variable commuting with the
Hamiltonian of a {TLS}.

In the present note we propose an alternative approach which
takes into account the operator nature of different variables in
a more consistent way. This allows us to show that a quantum solid with
incorporated TLSs can demonstrate the presence of a pronounced mass
deficit only when an external force changes fast enough, but not in a
stationary situation or at very low frequencies. For comparison, we also
discuss the situation when the tunnelling inside the TLSs is incoherent.

{\bf 2. A solid with a quantum two-level system.}
Consider a {solid} of total mass $M$ incorporating a TLS
in which a group of atoms (or a single atom) of mass $m$ can tunnel between
two localized positions shifted by vector ${\bf a}$ with respect to the
rest of the {solid}.
Below  we use the term ``{solid frame}" to designate the
part of the {solid} which does not participate in the process of tunnelling
and the term ``system" to describe the whole system consisting of the
{solid frame} and the TLS.

In the absence of any interaction between the TLS and other internal degrees
of freedom (for example, phonons) the quantum-mechanical Hamiltonian of such
a system can be written as
\begin{equation}                                                \label{H0}
\hat{H}_0=\frac{1}{2M}{{\bf P}^2}-\varepsilon\hat{\sigma}_3+J\hat{\sigma}_1\,,
\end{equation}
where ${\bf P}=-i\hbar\partial/\partial{\bf R}$ is the operator of
total momentum of the system (conjugate to the center of mass
position ${\bf R}$), the second term describes the difference in energy
(given by $2\varepsilon$)
between the two localized states of the TLS
and the third one the process of quantum tunnelling
(with \makebox{amplitude $J$)} between these two states,
$\hat{\sigma}_1$ and $\hat{\sigma}_3$ being the Pauli matrices.

Naturally, the velocity of the center of mass of the system,
\begin{equation}                                               
{\bf V}\equiv\frac{d}{dt}{\bf R}=\frac{i}{\hbar}\left[{H}_0,{\bf R}\right]
       =\frac{1}{M}{\bf P}\,,
\end{equation}
is determined just by its total momentum ${\bf P}$ and is
insensitive to its internal life (a particular state of the TLS).
In the presence of an external force ${\bf F}(t)$ applied to the {system} as
a whole [which corresponds to replacing $\hat H_0$ by
$\hat H=\hat H_0-{\bf RF}(t)$], the time evolution of $\langle{\bf V}\rangle$
is entirely determined by relation 
$(d/dt){\bf V}={\bf F}/M$.

However, in a number of experimental situations an external force
(for example, of mechanical origin) is applied not to the center of mass
of the {system}, but to the {solid frame}, and one is interested in the
relation between this force ${\bf f}(t)$ and the velocity of
the {solid frame}\footnote{Andreev calls the same variable
``the solid bulk velocity" \cite{A07,A08}.},
${\bf v}(t)$, which is not obliged to have exactly
the same form as the relation between ${\bf V}$ and ${\bf F}$.
The coordinate describing the position of the {solid frame}, ${\bf r}$,
can be introduced by rewriting the definition of the center
of mass position ${\bf R}$ as
\begin{equation}                                              \label{M(X)}
M{\bf R}=(M-m){\bf r}+m{\bf x}=M{\bf r}+m({\bf x}-{\bf r})\,,
\end{equation}
where ${\bf x}$ is the position of the center of mass of the atoms
(atom) forming the TLS. After replacing ${\bf u}\equiv{\bf x}-{\bf r}$,
the displacement of the TLS with respect to the solid frame, by
$-\frac{1}{2}{\bf a}\hat{\sigma}_3$,
Eq. (\ref{M(X)}) can be rewritten as
\begin{equation}                                              
{\bf r}={\bf R}+\frac{m}{2M}{\bf a}\hat{\sigma}_3\,.
\end{equation}
The operator of the {solid frame} velocity
\makebox{${\bf v}\equiv(d/dt){\bf r}$} is then given by
\begin{equation}                                              \label{v}
{\bf v}=\frac{i}{\hbar}\left[{\hat H}_0,{\bf r}\right]
=\frac{1}{M}\left({\bf P}+\frac{mJ}{\hbar}{\bf a}\hat{\sigma}_2\right)\,.
\end{equation}

Naturally, this equation can be also rewritten as
\begin{equation}                                                \label{Q}
{\bf P}_{} = M{\bf v}-\frac{mJ{\bf a}}{\hbar}\,\hat\sigma_2\,,
\end{equation}
which corresponds to splitting the total momentum of the system ${\bf P}$
into the two terms related respectively with the {solid frame} and the
TLS. However it is important that in contrast to the center of mass
velocity ${\bf V}\equiv {\bf P}/M$, which in the absence of external force
commutes with the Hamiltonian, the velocity of the {solid frame} ${\bf
v}$ is an operator which does not commute with $\hat{H}_0$.

In the presence of an external force ${\bf f}(t)$ applied
to the {solid frame} the Hamiltonian of the system acquires form
\begin{equation}                                             
\hat H=\hat H_0-{\bf r}{\bf f}(t)
      =-{\bf R}{\bf f}(t)-h_a(t)\hat\sigma_\alpha\,
\end{equation}
where
\begin{equation}                                             
h_\alpha(t)=\left[-J,\;0,\;\varepsilon+\frac{m}{2M}{\bf a}{\bf f}(t)\right]
\end{equation}
plays the role of the effective magnetic field acting on the spin $1/2$
which can be associated with the TLS and subscript $\alpha=1,2,3$ denotes
the components of ${\bf h}$.

It follows from Eq. (\ref{v}) that
\begin{equation}                                            \label{dv/dt}
\frac{d}{dt}{\langle\bf v\rangle}
=\frac{{\bf f}}{M} +\frac{m}{M}\frac{J{\bf a}}{\hbar}
\frac{d}{dt}{\langle\sigma_2\rangle}\,,
\end{equation}
which suggests that the coefficient of proportionality between
$(d/dt)\langle{\bf v}\rangle$ and ${\bf f}$ can be different from $1/M$.
To find a more explicit form of a TLS-induced correction to the
{solid frame}'s equation of motion one needs to express the time derivative
of $\langle\sigma_2\rangle$ \makebox{in terms of ${\bf f}(t)$.}

{\bf 3. Adiabatic regime.}
If external force ${\bf f}(t)$ does not depend on time or changes very
slowly one needs to take into account the relaxation processes which force
vector $\langle\sigma_\alpha\rangle$ to remain always parallel to
$h_\alpha$, from where $\langle \sigma_2\rangle=0$. Therefore, when ${\bf
f}(t)$ changes sufficiently slowly the processes inside the TLS make
no corrections to ${\bf P}=M{\bf v}$.
Thus, in the adiabatic regime the presence of the TLS cannot lead to
the appearance of any difference between the real mass of the {system} and
its effective mass observed in experiments involving the application of
mechanical force.

The finite expression for the mass deficit in the stationary regime
derived by Andreev \cite{A07} in the easiest way can be reproduced by
calculating the average of the TLS contribution to the total momentum
${\bf P}$, that is of the second term in Eq. (\ref{Q}),
with the help of the Hamiltonian,
\begin{equation}                                                \label{HA}
\hat H_{} = \frac{M}{2}{\bf v}^2
-\frac{mJ}{\hbar}({\bf av})\,\hat\sigma_2
             -\varepsilon\hat\sigma_3+J\hat\sigma_1\,,
\end{equation}
which in Ref. \cite{A07} is obtained by applying to \makebox{$H_{\rm
TLS}=-\varepsilon\hat\sigma_3+J\hat\sigma_1$} the Galilean transformation
from the reference frame in which the {solid frame} is at rest to the
reference frame moving with velocity ${\bf v}$. Naturally, the application
of this procedure implies that solid frame velocity ${\bf v}$ can be
treated as a classical variable.

One can easily check that Eq. (\ref{HA}) differs from Eq.
(\ref{H0}) with ${\bf P}$ replaced by
\makebox{$M{\bf v}-({mJ{\bf a}}/{\hbar})\hat\sigma_2$} only by a trivial
constant term which additionally tends to zero in the thermodynamic limit,
$m/M\rightarrow 0$. Thus the basic difference between the two approaches
consists only in choosing whether ${\bf P}$ or ${\bf v}$ is a classical
variable commuting with the Hamiltonian. The first option (adopted here)
leads to $\langle\hat{\sigma}_2\rangle=0$ and $\langle{\bf P}\rangle=M{\bf
v}$, whereas the second one produces for the average of the TLS
contribution to the total momentum
the expression \cite{A07}
\begin{equation}                                              
\langle{\bf P}_{\rm TLS}\rangle=-\left(\frac{mJ}{\hbar}\right)^2
\frac{\tanh[E({\bf v})/T]}{E({\bf v})}{\bf a(av)}
\end{equation}
with
\begin{equation}                                              
E({\bf v})\equiv\sqrt {\varepsilon^2+J^2+\left({mJ}/{\hbar}\right)^2({\bf
av})^2}\,,
\end{equation}
which does not respect the Galilean invariance being nonlinear in ${\bf
v}$.

{\bf 4. Finite-frequency response.}
When interaction of the TLS with other degrees of freedom
(heat bath) is weak, the basic form of the linear response of the TLS
to the application of the external force with a finite frequency,
${\bf f}(t)\propto\cos(\omega t)$, can be found by constructing
the periodic solution of the equations describing the free evolution of
the TLS \cite{A08},
\begin{equation}                                            \label{ds/dt}
\frac{d}{dt}\hat{\sigma}_\alpha
=\frac{i}{\hbar}\left[H,\hat{\sigma}_\alpha\right]
=-\frac{2}{\hbar}\epsilon_{\alpha\beta\gamma}h_\beta\hat{\sigma}_\gamma\,.
\end{equation}
After diagonalizing $H_{\rm TLS}=-\varepsilon\hat\sigma_3+J\hat\sigma_1$
and constructing the corresponding finite-temperature density matrix
$\hat{\rho}=\exp\left(-\hat{H}_{\rm TLS}/T\right)$ one obtains 
that in the absence of an external force
\begin{equation}                                            \label{sigma0}
\langle\hat{\sigma}_\alpha\rangle^{(0)}
=\left(-\frac{J}{E},~0,~\frac{\varepsilon}{E}\right)\tanh\frac{E}{T}\;
\end{equation}
with $E\equiv\sqrt{\varepsilon^2+J^2}$.

Solution of the equations for $\langle\hat{\sigma}_\alpha\rangle$ obtained
by the linearization in the vicinity of
$\langle\hat{\sigma}_\alpha\rangle^{}
=\langle\hat{\sigma}_\alpha\rangle^{(0)}$ gives
\begin{equation}                                             \label{dsy/dt}
\frac{d}{dt}{\langle\hat{\sigma}_2\rangle}
=-\frac{m}{M}\frac{\omega^2}{\omega^2-\Omega^2}
\frac{{\bf a}{\bf f}(t)}{\hbar}
\langle\hat{\sigma}_1\rangle^{(0)}\,
\end{equation}
with $\Omega=2E/\hbar$. Substitution of Eq. (\ref{sigma0}) into Eq.
(\ref{dsy/dt}) and then into Eq. (\ref{dv/dt}) transforms the latter into
\begin{equation}                                           \label{dv/dt-b}
\frac{d}{dt}{\langle v_i\rangle}=\left[\frac{\delta_{ij}}{M}
+\frac{\omega^2}{\omega^2-\Omega^2}
\frac{ {a}_i {a}_j}{M^2}\lambda
\right]{f}_j(t)\,,
\end{equation}
where subscripts $i$ and $j$ denote the components of vectors
in the real space and
\begin{equation}                                          
    \lambda\equiv\lambda(T)=\frac{m^2J^2}{\hbar^2 E}\tanh\frac{E}{T}\;.
\end{equation}

When one additionally takes into account the processes of transverse
relaxation induced by the interaction with the heat bath\footnote{The
interaction with the heat bath can also lead to the renormalization of $J$
\cite{LCDFGZ}.}, the
singularity at $\omega=\Omega$ is smeared out and the TLS-induced
correction to $(d/dt)\langle v_i\rangle$ acquires also a dissipative
contribution, proportional not to $\cos(\omega t)$ but to $\sin(\omega
t)$. If for simplicity one assumes that the relaxation can be
characterized by the same relaxation time $\tau$ (with $\Omega\tau\gg 1$)
for both directions perpendicular to
$\langle\hat{\sigma}_\alpha\rangle^{(0)}$,
the form of the result corresponds to the replacement of
$\omega^2-\Omega^2$ in the denominator in Eq. (\ref{dv/dt-b}) by
$(\omega+i/\tau)^2-\Omega^2$. The dissipative contribution
is the dominant one when $\tau|\omega-\Omega|\ll 1$.

When the system contains many TLSs with random orientations, Eq.
(\ref{dv/dt-b}) can be replaced by \makebox{$(d/dt)\langle{\bf
v}\rangle={\bf f}(t)/M_{\rm eff}(\omega)$} with
\vspace*{-2mm}
\begin{equation}                                             \label{1/M*}
    \frac{1}{M_{\rm eff}(\omega)}=\frac{1}{M}+\frac{1}{3M^2}
    \sum_n\frac{\omega^2}{(\omega+i/\tau_n)^2-\Omega_n^2}\gamma_n a_n^2\;,
\end{equation}
where subscript $n$ numbers different two-level systems and we have
assumed that vectors ${\bf a}_n$ have an isotropic distribution.
As usual with such a notation, the real and imaginary parts of
$1/M_{\rm eff}(\omega)$ describe the amplitudes of terms in
$(d/dt)\langle{\bf v}\rangle$ which are proportional  respectively to
$\cos(\omega t)$ and $\sin(\omega t)$ when an external force applied
to the solid frame is proportional to $\cos(\omega t)$.

It follows from the structure of Eq. (\ref{1/M*}) that when $\omega$ is
much larger than all $\Omega_n$ the form of $M_{\rm eff}(\omega)$ indeed
corresponds to the presence of frequency-independent mass deficit (as
it was suggested by Andreev \cite{A08}). With the decrease in frequency
the value of mass deficit decreases and passes through zero at frequencies
at which the dissipative contribution is most prominent. At
$\omega\rightarrow 0$ both parts (dissipative and nondissipative) of the
TLS-induced correction to $1/M$ tend to zero.

{\bf 5. The case of incoherent tunnelling.\label{IT}}
Consider now the case when the tunnelling process inside a two-level
system is incoherent. 
In such a situation the displacement of the TLS with respect to
the solid frame ${\bf u}\equiv {\bf x}-{\bf r}$ induced by a
time-dependent external force ${\bf f}(t)\propto\cos(\omega t)$ coupled
to the solid frame's position ${\bf r}={\bf R}-(m/M){\bf u}$
acquires a very simple form \cite{KV},
\begin{equation}                                                 \label{u}
    {\bf u}(\omega)=-\frac{1}{-i\tau\omega+1}\frac{m}{M}
    \frac{{\bf a}({\bf af})}{4T\cosh^2(\epsilon/T)}
\end{equation}
where relaxation time $\tau$ is inversely proportional to the tunnelling
rate, whereas $\varepsilon$ retains the same meaning as above.

Substitution of Eq. (\ref{u}) into the classical analog of Eq.
(\ref{dv/dt}), namely
\begin{equation}                                                 
\frac{d}{dt}{\bf v}
=\frac{1}{M}\left[{\bf f}-\frac{m}{M}\frac{d}{dt}\frac{d\bf
u}{dt}\right]\,,
\end{equation}
then leads to the following expression for the frequency-dependent
coefficient of proportionality in the relation
$(d/dt){\bf v}={\bf f}/M_{\rm eff}$:
\begin{equation}                                            \label{1/Meff}
\frac{1}{M_{\rm eff}(\omega)}=\frac{1}{M}-\frac{\omega^2}{12TM^2}
\sum_n
\frac{1}{-i\tau_n\omega+1}\frac{m_n^2a_n^2}{\cosh^2(\varepsilon_n/T)}
\end{equation}
The form of Eq. (\ref{1/Meff}) demonstrates that in the case of incoherent
tunnelling the TLSs-induced contribution to the effective mass can never be
described in the form of frequency independent mass deficit and at high
frequencies is of the dissipative nature.

\addtolength{\textheight}{-17mm}

{\bf 6. Conclusion.}
In the present note we have analyzed how the presence of TLSs influences
the dynamic properties of a solid and have demonstrated that in order to
observe a reduction of the effective mass of the sample (analogous to that
in superfluids) the solid should incorporate quantum TLSs and the
frequency at which the external force is applied should be high enough in
comparison with resonance times of the TLSs. In the case
of incoherent tunnelling the regime in which the contribution from the
TLSs can be described as frequency-independent effective mass deficit is
absent. Since TLSs are local objects, analogous conclusions are applicable
also to the effective moment of inertia $I_{\rm eff}(\omega)$, which is
a relevant quantity in torsional oscillator experiments. Experimentally,
the crossover to the regime which can be associated with the appearance
of an effective mass deficit has been demonstrated in a number
of works starting from those of Kim and Chan \cite{KCh}.
The detailed comparison of the temperature dependences of both components
of the response can be found in Ref. \cite{HPG}, which also contains a
comprehensive list of references to other experimental and theoretical works.

An essential feature of our results is that in both regimes
(of coherent and incoherent tunnelling) the TLS-induced contribution
vanishes in the limit of \makebox{$\omega\rightarrow 0$}.
Another common feature of the two cases is that
the presence of the TLSs makes an additive
contribution to $1/M_{\rm eff}$ or $1/I_{\rm eff}$ rather than to $M_{\rm
eff}$ or $I_{\rm eff}$. In that respect the situation is quite analogous
to that in superconducting vortex glasses where TLSs make a positive
contribution to the inverse superfluid density (in other terms, specific
inductance) rather then a negative contribution to the superfluid density
itself \cite{KV,K01}.

It seems worthwhile to mention that the phenomenological approach
\cite{NBGT} used in Ref. \cite{HPG} for analyzing the experimental data is
based on conjectures which are in contradiction with both these
properties. Namely, the authors of Ref. 
\cite{NBGT} have assumed that the internal degrees of freedom of solid
$^4$He make a contribution to its back action which in terms of the
frequency-dependent effective moment of inertia of $^4$He sample, $I_{\rm
eff}(\omega)$, can be written as an additive correction to the
frequency-independent moment of inertia of the solid frame,
\begin{equation}                                              \label{I}
I_{\rm eff}(\omega)=I_0+\frac{g_0}{\omega^2(1-i\tau\omega)^\beta}\;~~~
\end{equation}
with $\beta\leq 1$. Moreover, at low frequencies this contribution
does not tend to zero but behaves itself like a {\em negative}
correction to the stiffness constant of the torsional oscillator.
If it were really so, the solid by itself
(outside of the torsional oscillator) would be unstable.
A more logical assumption on the form of $I_{\rm eff}(\omega)$
in a glassy system with incoherent tunnelling inside the TLSs
[consistent with the form of Eq. (\ref{1/Meff})] would be
\begin{equation}                                             \label{I-2}
I_{\rm eff}(\omega)=\left[I^{-1}_0-\frac{{\bar g}\omega^2}
{(1-i\tau\omega)^\beta}\right]^{-1}.~~~
\end{equation}

The main argument of Ref. \cite{HPG} in favor of the superglass state
consists in the impossibility to reconcile experimental data with
dependence (\ref{I}) - the observed frequency shift is too large in
comparison with
{the maximum} in dissipation
(characterized by the inverse quality factor), the same being true
also for dependence (\ref{I-2}).
However, the dispersion  of the parameters of the TLSs can be taken into
account by the replacement of $\beta=1$ by $\beta<1$ only for a particular
form of the distribution and in other cases may lead to different
dependences, especially if the appearance of the frequency shift is
induced by the crossover between the regimes of incoherent
and coherent tunnelling inside the TLSs.
The temperature dependence of $\bar g$, the amplitude of the TLS-induced
term, also may change the relation between the frequency shift and
the minimal quality factor. To clarify the situation,
the experimental investigations of the temperature dependence of the
solid $^4$He dynamic response have to be complemented
by more systematic studies of its frequency dependence.

\vspace*{3mm}

The author is grateful to A.\,F. Andreev, { A.\,V. Balatsky}, Yu. Makhlin
and G.\,E. Volovik for useful discussions and comments. This work has been
supported by the RFBR Grant No. \makebox{09-02-01192-a} and by the RF
President Grant for Scientific Schools No. 5786.2008.2.

\end{document}